\documentclass[5p,twocolumn,numbers,sort]{elsarticle}

\usepackage{url}
\usepackage{amsmath}
\usepackage{booktabs}
\usepackage{balance}
\usepackage{graphicx}
\usepackage{tabularx}
\usepackage[colorlinks = true,
  linkcolor = blue,
  citecolor = blue,
  urlcolor = blue]{hyperref}

\usepackage{fancyhdr} % for arxiv
\fancypagestyle{firststyle}
{
   \fancyhf{}
   \fancyfoot[C]{\scriptsize{Journal of Information Security and Applications, 2025, vol.~90, pp.~103989. This version is the author's copy. \\ The publisher's definite version is available at Elsevier via \url{https://doi.org/10.1016/j.jisa.2025.103989}.}}
}

\begin{document}

\title{SoK: The Design Paradigm of Safe and Secure Defaults}

\author[sdu]{Jukka Ruohonen\corref{cor}}
\ead{juk@mmmi.sdu.dk}
\address[sdu]{University of Southern Denmark, S\o{}nderborg, Denmark}

\begin{abstract}
In security engineering, including software security engineering, there is a
well-known design paradigm telling to prefer safe and secure defaults. The paper
presents a systematization of knowledge (SoK) of this paradigm by the means of a
systematic mapping study and a scoping review of relevant literature. According
to the mapping and review, the paradigm has been extensively discussed, used,
and developed further since the late 1990s. Partially driven by the insecurity
of the Internet of things, the volume of publications has accelerated from the
circa mid-2010s onward. The publications reviewed indicate that the paradigm has
been adopted in numerous different contexts. It has also been expanded with
security design principles not originally considered when the paradigm was
initiated in the mid-1970s. Among the newer principles are an ``off by default''
principle, various overriding and fallback principles, as well as those related
to the zero trust model. The review also indicates problems developers and
others have faced with the~paradigm.
\end{abstract}

\begin{keyword}
fail-safe defaults; security engineering; security design principles; scoping
review; systematic mapping study
\end{keyword}

\maketitle

\section{Introduction}

\thispagestyle{firststyle} % for arxiv

The design paradigm of safe and secure defaults traces to the classical work by
Saltzer and Schroeder who derived eight general principles for computer,
software, and cyber security in general~\cite{Saltzer75}. These are shown in
Table~\ref{tab: saltzer}. In particular, the paradigm's roots originate from the
principle of fail-safe defaults. This principle is about access controls; a
default should be a lack of access, and a design and its implementation should
identify the particular conditions upon which access can be permitted.

Upon replacing the noun access with some other suitable word, the principle
generalizes to a broader notion that defaults should be safe and secure. With
this generalization, also many of the other design principles in Table~\ref{tab:
  saltzer} are related to the fail-safe defaults. For instance, it could be
argued that a system's or a software's defaults should satisfy the principle of
separation of privileges; in modern terms, it would mean that two-factor or
multi-factor authentication would be a default.

\begin{table}[th!b]
\centering
\caption{The Eight Security Design Principles of Saltzer and Schroeder \cite[quotations from pp.~1282--1283]{Saltzer75}}
\label{tab: saltzer}
\begin{small}
\begin{tabularx}{\linewidth}{lX}
\toprule
Principle & Quotation \\
\hline
Economy & ``Keep the design as simple and small as possible.'' \\
\cmidrule{2-2}
Fail-safe defaults & ``Base access decisions on permission rather than exclusion.'' \\
\cmidrule{2-2}
Complete mediation & ``Every access to every object must be checked for authority.'' \\
\cmidrule{2-2}
Open design & ``The design should not be secret.'' \\
\cmidrule{2-2}
Separation of privilege & ``Where feasible, a protection mechanism that requires two keys to unlock it is more robust and flexible than one that allows access to the presenter of only a single key.'' \\
\cmidrule{2-2}
Least privilege & ``Every program and every user of the system should operate using the least set of privileges necessary to complete the job.'' \\
\cmidrule{2-2}
Least common mechanism & ``Minimize the amount of mechanism common to more than one user and depend on by all users.'' \\
\cmidrule{2-2}
Psychological acceptability & ``It is essential that the human interface be designed for ease of use, so that users routinely and automatically apply the protection mechanisms correctly.'' \\
\hline
\end{tabularx}
\end{small}
\end{table}

The Saltzer's and Schroeder's paper is a modern classic in computer
science. Yet, thus far, no systematic reviews have been conducted about its use
in academic research, although there are some existing reviews that come
close~\cite{vandenBerghe17}. The review presented fills the apparent knowledge
gap. In addition, the review fulfills other common functions of literature
reviews, including the identification of trends and patterns, clarification of
complex concepts, and support for education through cataloging important or even
foundational knowledge~\cite{Kraus24}. Regarding trends, as will be seen, the
design paradigm has been continuously discussed in the face of new
technologies. It has also been extended and customized to meet requirements
originating from new design problems. Furthermore, it has frequently been
misused in practice. This misuse raises also the review's practical
relevance. As will be pointed out later on in the concluding Section~\ref{sec:
  discussion}, secure defaults are mandated by recent cyber security
regulations. Thus, the SoK is beneficial also for practitioners and regulators
who both may find new insights from the knowledge cataloged.

A couple of terminological clarifications are needed before continuing. The
first clarification is that the topic is about design \textit{principles}, which
are broader and more theoretical than design \textit{patterns}. An analogy from
programming suffices to elaborate the distinction; reusability is a common
software design principle, whereas design patterns often allow to satisfy this
principle through a customized use of recurring design solutions to common
design \textit{problems}~\cite{Gamma95}. Then, in the present context the
overall problem is about insecurity, unsafety, or both. The second clarification
follows; the distinction between safety and security is debated and
blurry~\cite{Ruohonen22ICLR} . In what follows, unsafety is understood to refer
to unintentional mistakes often originating from poor designs that have
consequences particularly for the health and well-being of humans. In contrast,
insecurity is taken to involve a potential of intentional attacks or other
disturbances against a design and its implementation. Note that insecurity may
cause unsafety, but the reverse relation is often less clear.

The focus on security design principles frames the paper toward security
engineering. Alas, there is no single universally agreed definition for security
engineering. In general, it can be seen to be about putting ``security theory
into security practice'', meaning that ``a security engineer designs and makes
systems that are protected against threats, i.e. forces to which systems may be
subjected'' \cite[p.~59]{Schumacher03}. In the present context the notion about
putting theory into practice is about adapting and applying the abstract design
paradigm in more practical contexts. The framing toward security engineering is
also important because it puts the paradigm into a context of engineering secure
systems; hence, the research reviewed is also largely about defensive cyber
security. Although there are papers about vulnerabilities in existing systems,
whether hardware or software, the majority of papers reviewed are about building
or improving systems, software, networks, protocols, or technologies in general
with security and safety in mind.

% https://www.sciencedirect.com/science/article/abs/pii/S0164121223000493
% "allowing practitioners to discover high-quality, evidence-based approaches to harness the power of human aspects in software engineering"

As for the paper's remaining structure, the literature reviewing methodology is
elaborated in the opening Section~\ref{sec: methodology}. The actual review is
presented in the subsequent Section~\ref{sec: review}. A conclusion and an
accompanying discussion are presented in the final Section~\ref{sec: discussion}.

\section{Methodology}\label{sec: methodology}

The paper is a systematic mapping study---or, alternatively, a scoping
review. Both systematic mapping studies and scoping reviews share the same
overall rationale. Their use is often justified when operating in a heterogenous
context in which traditional empirical evidence may be lacking and multiple
disciplines may operate. In addition, both literature reviewing techniques are
often justified when feasibility is a concern; the techniques are useful for
determining a value as well a potential scope and a cost of undertaking a
full-blown systematic literature review~\cite{Haddaway16, Pham14}. If systematic
literature reviews are seen to align with confirmatory \text{research---after}
all, they try to collate all evidence about a topic, systematic mapping studies
and scoping reviews are more on the exploratory side of things. In general, they
explore the nature and scope of existing literature, trying to inform future
research, including further reviews, by identifying not necessarily evidence
gaps but knowledge gaps in general~\cite{Peters21}. These general
characterizations justify also the paper's reviewing methodology; as will be
seen, the topic is highly heterogenous without much traditional empirical
evidence.

That said, both scoping reviews and systematic mapping studies have adopted from
systematic literature reviews the good practice of using a structured and
transparent protocol for searching literature. Thus, the literature search
protocol used in the present review is:
\begin{align*}
(\textmd{safe}~\textmd{AND}~\textmd{default})~\textmd{OR}~
(\textmd{secure}~\textmd{AND}~\textmd{default}) ,
\end{align*}
where AND and OR are Boolean operators. The two clauses separated by OR capture
both security engineering design practices and more general design solutions
involving safe defaults. They also capture common associated phrasings such as
fail-safe defaults.

In addition to removing duplicates, four criteria were used to exclude less
relevant or otherwise ill-suited literature. The first criterion was that only
primary studies were included; hence, a paper merely citing in passing some
other paper that discusses safe or secure defaults did not qualify. The second
criterion was that a qualifying paper had to discuss safe or secure defaults by
providing a definition, an example, a rationale, an explanation, or some related
elaboration. Thus, those papers were excluded that simply pointed out that there
is a design paradigm involving safe and secure defaults. The same applies to
drive-by mentions of general design ideals, such as ``secure by default'',
``privacy by default'', and ``security by design''. Likewise, many papers
discussing default values for algorithms were excluded. The third criterion
imposed was that only peer reviewed journal articles and publications in
conference proceedings were qualified; hence, book chapters, standards,
editorials, and related content were excluded. The fourth criterion was simple:
all papers addressing financial matters were excluded; therein, the word default
has a different~meaning.

\begin{figure}[th!b]
\centering
\includegraphics[width=\linewidth, height=4.7cm]{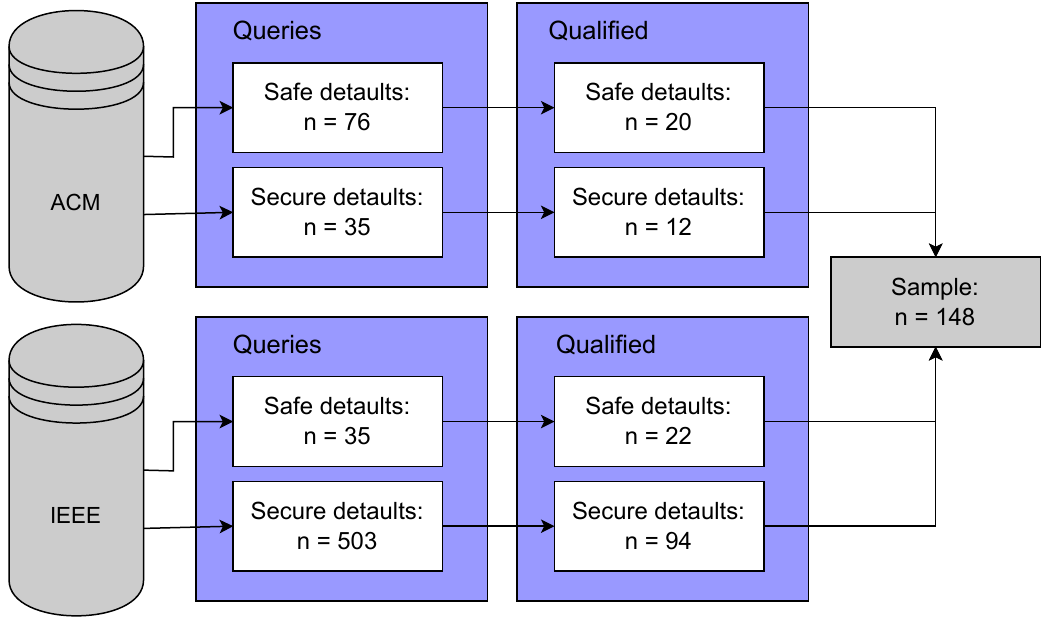}
\caption{The Sample of Literature Reviewed}
\label{fig: sample}
\end{figure}

Even with the criteria, preliminary queries indicated a vast amount of peer
reviewed literature. To deal with this severe feasibility obstacle, the final
searches were restricted only to the ACM's and IEEE's electronic libraries. This
restriction underlines the paper's nature as a \textit{scoping}
review. Regarding the noted value contemplation before conducting a full
systematic literature review, it can be tentatively concluded that not much
additional value would be supposedly available because the $n = 148$ reviewed
papers (see~Fig.~\ref{fig: sample}) are already sufficient for conveying the
relevant points raised and soon~discussed. If further databases would be
queried, it also remains unclear whether it would be possible to review the
literature without resorting to natural language processing or
bibliometrics.

It can be also noted that the criteria were strictly followed in order to
minimize subjectivity. Therefore, it is worth further remarking that there is a
paper~\cite{ieee71secure} that has likely used so-called tortured
phrases~\cite{Cabanac21} or something alike. To deal with such papers as well as
the overall feasibility problem, it might be possible to evaluate the quality of
papers~\cite{Zolduoarrati23}, but the problem is that there is no
universally agreed definition for paper quality~\cite{Brereton07}. This lack of
a definition would presumably increase subjectivity. A~potential solution would
be to focus on national rankings of publication venues~\cite{Galster14}, journal
impact factors, or related quantitative information, but also this choice would
likely bias a literature sample because high-quality work is published also in
low-prestige venues, and the other way around. These points notwithstanding, it
can be remarked that the overall quality of the sample is good based on a
subjective evaluation. This remark can be reinforced by taking a look at the
publication venues; many, but not obviously all, papers were published in
well-known or even top-ranked security conferences and journals.

Finally, the reviewing itself follows a thematic approach that is common in
qualitative research~\cite{Ruohonen24FM, Ruohonen24JSS}. In other words, the
goal is to capture and categorize the major themes present in the
literature. The thematic approach is also necessary due to the large amount of
literature sampled. The themes were constructed by first identifying a large
number of distinct categories, which were then narrowed and sharpened by
collating and merging of overlapping, overcasting, or otherwise fuzzy categories
~\cite{Ruohonen24JSS, Williams19}. In practice, the identification of themes and
the reviewing in general were started by reading each paper's abstract. The
second step was about reading the portions in which defaults were discussed. If
a more detailed reading was required to deduce about the defaults, a paper's
introduction and conclusions were first read. If these were not sufficient, as
was the case with a few papers, a full reading was conducted, from the first to
the last sentence.

\section{Review}\label{sec: review}

In what follows, the review is presented in five subsections. The first
subsection presents a few quantitative observations about the literature
sample. The two subsequent subsections elaborate the computing domains in which
the papers have operated together with their motivations. The design principles
behind the paradigm are discussed in the next subsection. The final subsection
briefly discusses some problems identified in the literature.

\subsection{Quantitative Observations}

\begin{figure}[th!b]
\centering
\includegraphics[width=8cm, height=3cm]{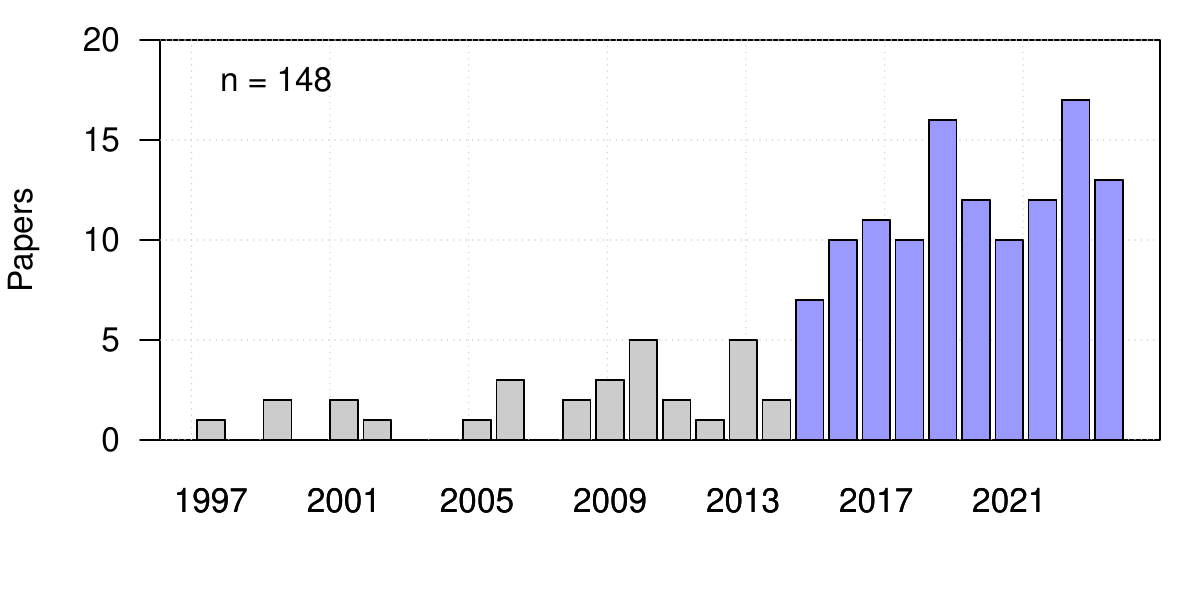}
\caption{Publication Years}
\label{fig: years}
\end{figure}

The classical paper of Saltzer and Schroeder \cite{Saltzer75} was published in
1975. To this end, a good way to start the review is to look at the publication
years of the papers sampled. These are shown in Fig.~\ref{fig: years}.

\begin{figure}[th!b]
\centering
\includegraphics[width=\linewidth, height=10cm]{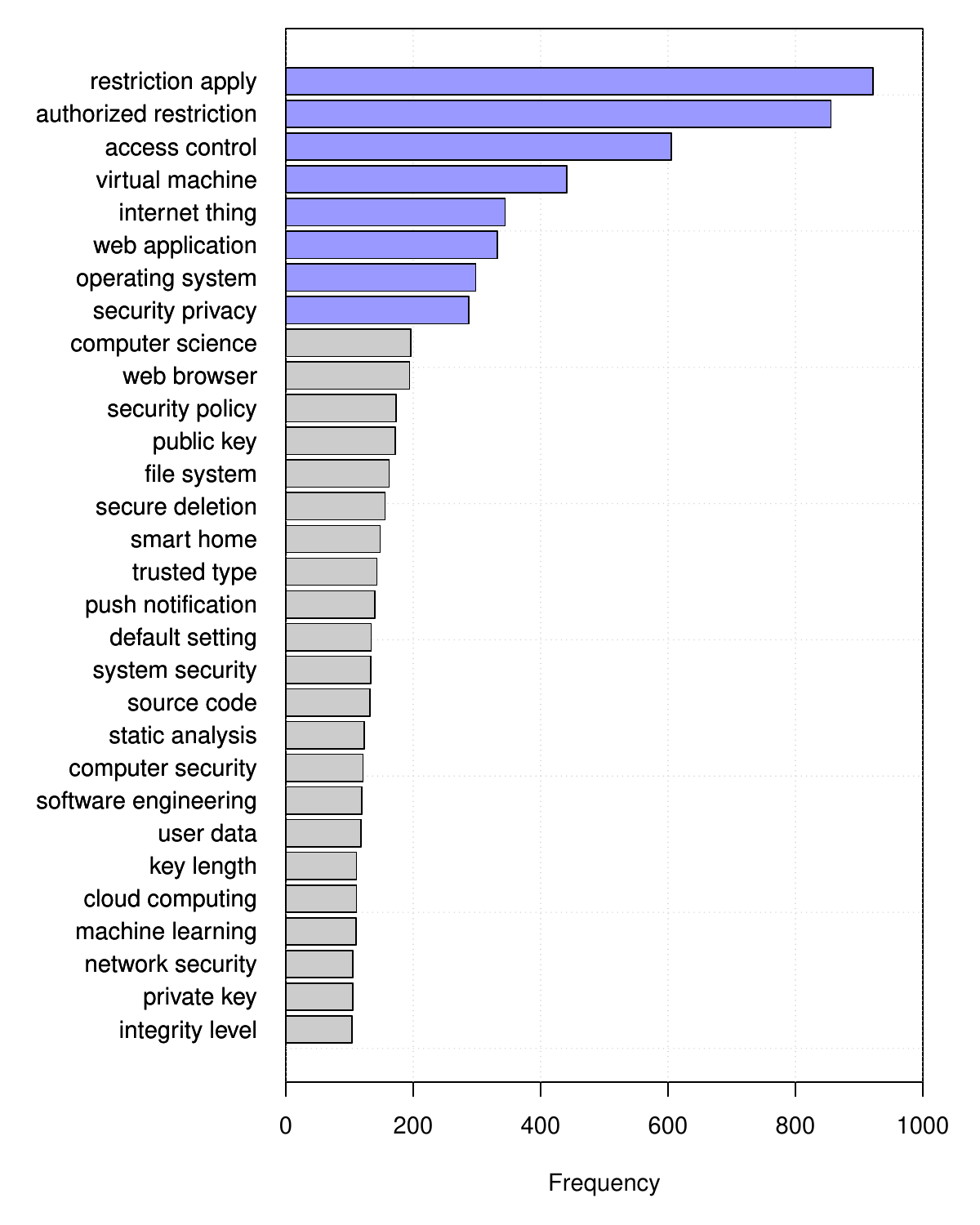}
\caption{Top-30 Bigrams}
\label{fig: bigrams}
\end{figure}

As can be seen, there has been a growing interest in the design
paradigm. However, the earliest papers were published in the late 1990s, which
marks an over twenty years gap between the classic and the initially following
contributions. A further point is that the publication pace has accelerated from
about mid-2010s onward. As soon discussed, there is a specific reason for
the~acceleration.

The classical paper has been explicitly cited only in two papers sampled,
meaning that the full title of the paper appears in bibliographies. Implicitly,
however, the surnames Saltzer and Schroeder appear together in about seven
percent of the $148$ papers sampled. Furthermore, either the character string
\texttt{fail-safe default} or the character string \texttt{fail safe default}
appears in about eight percent of the papers sampled; these strings were
searched from anywhere in the lower-cased textual representations of the
papers. Thus, it may be that the 1970s paper is not that well-known after
all---or it may be that it already belongs to the common pool of computing
knowledge, such that citing it is no longer necessary.

Further basic quantitative information can be provided in the form of the
top-ranked bigrams shown in Fig.~\ref{fig: bigrams}. These were constructed from
the lower-cased textual representations of the papers with a conventional
pre-processing strategy: the texts were tokenized according to white space and
punctuation characters; only alphabetical tokens recognized as English words
were included; the tokens were lemmatized into their basic dictionary forms;
only lemmatized tokens longer than three characters and shorter than twenty
characters were included; and stopwords were excluded.\footnote{~This
pre-processing strategy follows existing research from which further details can
be found~\cite{Ruohonen22IS}. In the present context it is only relevant listing
the custom stopwords used: \textit{article}, \textit{available},
\textit{conference}, \textit{downloaded}, \textit{fig}, \textit{figure},
\textit{future}, \textit{international}, \textit{licensed}, \textit{limitation},
\textit{online}, \textit{related}, \textit{restriction}, \textit{shown},
\textit{southern}, \textit{university}, \textit{use}, \textit{vol}, and
\textit{work}. All are related to boilerplate text that appears in publications,
some of which is added by publishers} Then, as can be seen from the figure,
already the most frequent bigrams indicate a presence of numerous distinct
contextual domains. These are disseminated~next.

\subsection{Domains}

A good way to continue with the review is to present the particular computing
domains in which the studies reviewed operated. These are enumerated in
Table~\ref{tab: domains}. As the review mixes also a few papers that do not
belong to the sample formally gathered, the table serves also as a bookkeeping
material about the literature explicitly reviewed. In what follows, a term
non-sampled literature is used to distinguish papers referenced that are not
part of the actual sample.

As can be seen from the table, safe and secure defaults have been discussed in a
number of distinct domains, ranging from traditional computer science and
software engineering domains, such as cryptography, databases, operating
systems, and programming languages to newer domains and technologies, such as
smarthomes and blockchains, and from there to a little more unconventional
domains, such as psychology and even work safety. Regarding the unconventional
domains, a couple of concrete examples can be given. The first is a paper
investigating secure deletion of files; it is noteworthy because data is not
generally deleted securely by default~\cite{ieee27secure}. Thus, the paper
demonstrates how fundamental the paradigm's violations can be. The same point
applies to the couple of papers dealing with hardware. The second outlying
example is a paper investigating the use of A/B testing for experimenting how
users react to new designs, features, or bug fixes. It is noteworthy because it
demonstrates how the design paradigm can emerge as a side-product; among the
paper's conclusions is a recommendation that developers should be educated to
conduct experiments in such a way that a safe default state is always maintained
and all leftovers are cleaned afterwards~\cite{acm4safe}. Although not captured
by the literature search protocol, the clean-up recommendation is also known as
a remnant removal design principle; a terminating system should clear past
traces that are not required for later use~\cite{vanOorschot21}. As may be
common to design principles in general, the remnant removal design principle
might be also known with some other term in some other context.

\begin{table*}[th!b]
\centering
\caption{Contextual Domains}
\label{tab: domains}
\begin{tabularx}{\linewidth}{lX}
\toprule
Domain & Papers \\
\hline
A/B testing & \cite{acm4safe} \\
Artificial intelligence & \cite{ieee71secure} \\
Blockchains & \cite{ieee17safe} \\
Configuration management & \cite{ieee7safe, ieee31secure} \\
Cyber-physical systems & \cite{acm10safe, ieee6safe, ieee10safe, ieee12safe, ieee18safe, ieee21safe, ieee77secure} \\
Cryptography & \cite{acm8safe, acm9safe, acm12safe, acm13safe, acm5secure, ieee63secure, ieee66secure, ieee84secure, ieee91secure} \\
Databases & \cite{acm21safe, ieee29secure} \\
Data deletion & \cite{ieee27secure} \\
Distributed systems & \cite{acm7safe, acm19safe} \\
Education & \cite{acm3safe} \\
Emails & \cite{ieee11safe, ieee36secure} \\
Embedded devices (including~IoT) & \cite{acm5safe, acm1secure, ieee9safe, ieee3secure, ieee6secure, ieee9secure, ieee18secure, ieee21secure, ieee23secure, ieee37secure, ieee40secure, ieee41secure, ieee43secure, ieee45secure, ieee46secure, ieee51secure, ieee54secure, ieee55secure, ieee56secure, ieee58secure, ieee59secure, ieee60secure, ieee69secure, ieee70secure, ieee74secure, ieee82secure, ieee83secure, ieee90secure} \\
Governmental systems & \cite{acm18safe} \\
Hardware & \cite{ieee78secure, ieee81secure} \\
Healthcare & \cite{acm6safe, ieee10secure} \\
Honeypots & \cite{ieee26secure} \\
Networks (excluding~IoT and web) & \cite{acm11safe, acm7secure, ieee1safe, ieee4safe, ieee14safe, ieee16safe, ieee22safe, ieee8secure, ieee14secure, ieee16secure, ieee19secure, ieee24secure, ieee28secure, ieee30secure, ieee33secure, ieee35secure, ieee38secure, ieee39secure, ieee49secure, ieee57secure, ieee61secure, ieee68secure, ieee72secure}  \\
Operating systems (including virtualization) & \cite{acm20safe, ieee7secure, ieee11secure, ieee15secure, ieee25secure, ieee34secure, ieee44secure, ieee47secure, ieee48secure, ieee52secure, ieee53secure, ieee64secure, ieee65secure, ieee73secure, ieee75secure, ieee79secure, ieee87secure, ieee92secure} \\
Passwords & \cite{acm6secure, ieee4secure, ieee20secure, ieee32secure} \\
Programming and programming languages & \cite{acm15safe, acm16safe, ieee5secure, ieee22secure, ieee89secure} \\
Psychology & \cite{acm12secure} \\
Security engineering & \cite{acm2safe, acm4secure, ieee2safe, ieee19safe, ieee20safe, ieee88secure} \\
Smarthomes & \cite{acm11secure, ieee50secure, ieee80secure} \\
Smartphones & \cite{acm3secure, acm9secure, ieee2secure, ieee12secure, ieee62secure} \\
Social networks & \cite{ieee67secure, ieee93secure} \\
Trusted computing & \cite{ieee13secure} \\
Web & \cite{acm1safe, acm17safe, acm2secure, acm8secure, acm10secure, ieee3safe, ieee13safe, ieee15safe, ieee1secure, ieee17secure, ieee42secure, ieee85secure, ieee86secure, ieee94secure} \\
Work safety & \cite{ieee5safe, ieee8safe} \\
\bottomrule
\end{tabularx}
\end{table*}

Although a thematic analysis is not well-suited for quantitative insights, it
can be still concluded that embedded devices, including particularly the
Internet of things (IoT) devices, have been the leading force behind the design
paradigm---or, rather, its abuses. In fact, the bigram \texttt{internet thing}
takes the fifth place in Fig.~\ref{fig: bigrams}. This domain is also
referred to as the Internet of vulnerable things in the
literature~\cite{ieee18secure}. Alternatively, the insecurity of ``thingernets''
is related to the vulnerabilities in the ``thingabilities'' and
``thingertivity'' of things~\cite{ieee74secure}. To put the paper's humor aside,
the ``thingernets'' concept is illuminating in a sense that it has been
estimated that it was around the early 2010s when more ``thingabilities'' than
humans were connected to the Internet according to the non-sampled
materials~\cite{Cisco11}. Thus, the IoT domain largely explains also the
acceleration of papers discussing the design paradigm (cf.~Fig.~\ref{fig:
  years}). The reason is the domain's overall insecurity. In this regard, it is
worth continuing the historical narrative by pointing out a couple of papers
discussing the Mirai botnet built around compromised IoT
devices~\cite{ieee69secure, ieee83secure}. It was first discovered in 2016, and
rang the alarm bells throughout the computing world.

As will be soon discussed, the various vulnerabilities and insecurities plaguing
IoT devices are also a good example about software's role in the design paradigm
and its abuses. Regarding computer networks more generally, new paradigms, such
as software-defined networks (SDNs), further underline software's importance
also in domains not explicitly associated with software engineering. In any
case, the design principle has frequently been discussed and applied also in
other computer networking domains, including obviously the Internet but also the
world wide web and many other networking areas. Also operating systems have been
a frequent context in which the design principle has been discussed. In this
regard, it is worth pointing out that many of the early publications discussed
and used the paradigm in a context of access controls. Later on, the operating
system domain gained traction vis-\`a-vis the paradigm through the emergence of
virtualization and related techniques. These points can be drawn also from
Fig.~\ref{fig: bigrams} within which the bigrams \texttt{authorized
  restriction}, \texttt{access control}, \texttt{virtual machine}, and
\texttt{operating system} rank high. Of the newer contextual domains, it is
worth mentioning cyber-physical systems, including robotics, within which the
paradigm has often been discussed in relation to the safety of humans, machines,
and their interactions.

It is also worth noting a generic security engineering domain in Table~\ref{tab:
  domains}. The papers in this domain have addressed the security, usability,
documentation, and other related aspects of APIs, that is, application
programming interfaces~\cite{acm2safe}, developer-oriented software security
engineering~\cite{acm4secure}, using experiments to support security
designing~\cite{ieee88secure}, the design paradigm under
review~\cite{ieee2safe}, design of access controls~\cite{ieee19safe}, and the
practical use of the design principles behind the
paradigm~\cite{ieee20safe}. Particularly the two papers explicitly dealing with
and discussing the paradigm are worth emphasizing already because they also
signify the review's relevance. That is, there is an interest around the design
paradigm that goes beyond technical implementation~work.

\subsection{Motivations}

By and large, the whole paradigm has been motivated in the literature though a
negation. In other words, as can be also concluded from Fig.~\ref{fig:
  motivations}, which visualizes the motivating themes used in the papers, the
starting point and motivation have usually been unsafe and insecure
defaults. The motivations through a negation are expectedly particularly
pronounced in the IoT domain. Some publications talk about an insecure default
configuration problem, meaning that IoT devices are routinely shipped with
extremely poor configurations with respect to best---or even
\text{elementary---cyber} security practices~\cite{acm1secure}. As has been
pointed out in numerous publications, these configuration mistakes include, but
are not limited to, default or weak passwords. Regarding the other issues, the
IoT domain is shaped by ``extreme heterogeneity, mostly plug-and-play nature,
computational limitations, improper patch management, unnecessary open ports,
default or no security credentials, and extensive use of reusable open-source
software'' \cite[p.~11224]{ieee82secure}. On one hand: as with the Mirai botnet,
it is also worth emphasizing that the insecurity problems can escalate into
larger, even society-wide problems. Although many of the publications analyzed,
discussed, and elaborated the IoT domain with conventional consumer-grade
devices, such as routers and modems, the same issues seem to plague even smart
grid devices~\cite{ieee70secure}. On the other hand: although IoT devices have
likely exacerbated the problem, it is worth noting a publication from 2010 that
discussed the same issues with default passwords~\cite{ieee55secure}. Another
publication from 2012 analyzed the insecure default settings in Windows and
Linux operating systems at that time~\cite{ieee68secure}. Thus, like with the
design paradigm itself, the antonyms of unsafe and insecure defaults are nothing
new as such in the computing world.

\begin{figure*}[t!]
\centering
\includegraphics[width=12cm, height=10cm]{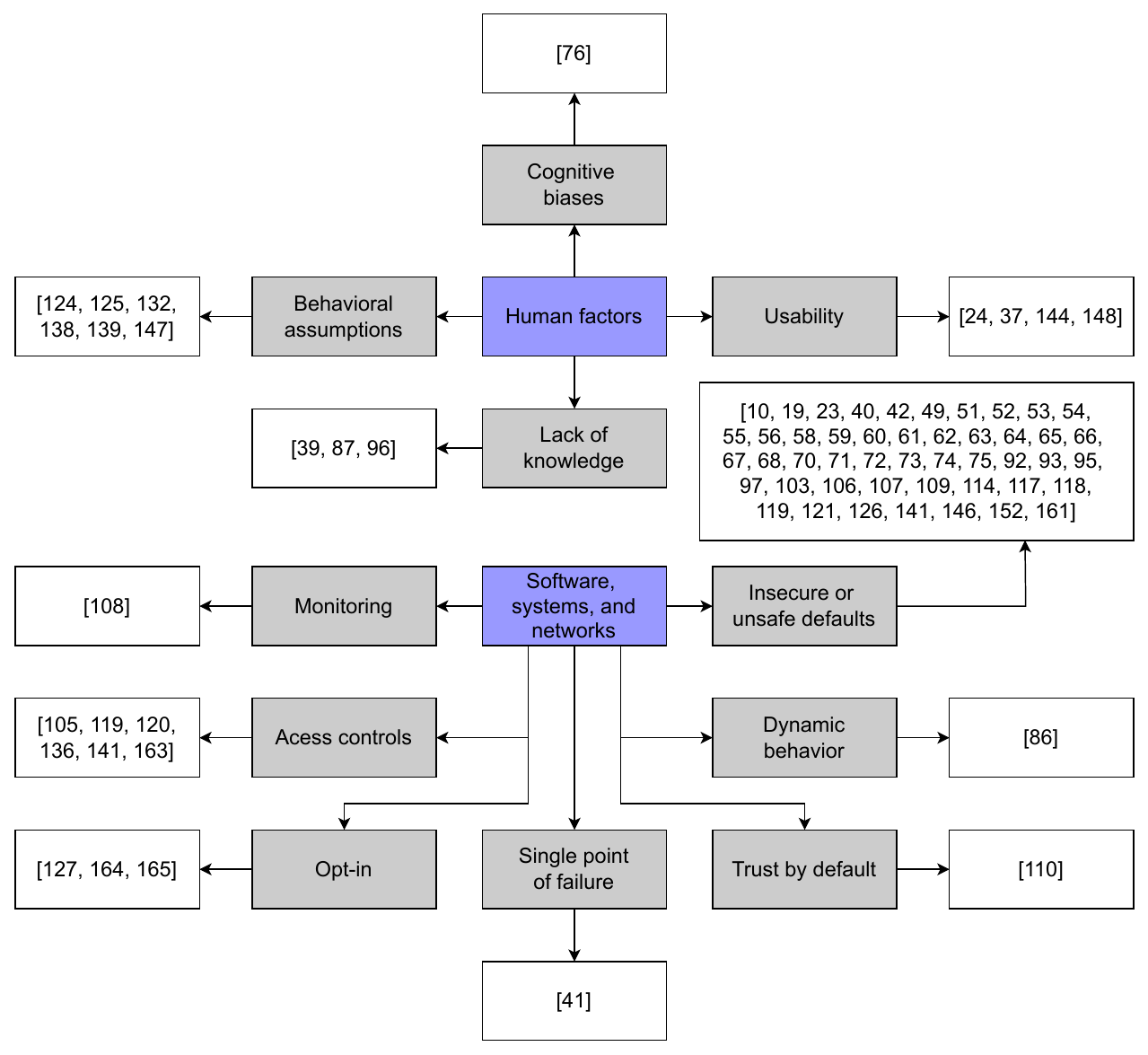}
\caption{Thematic Motivations for the Design Paradigm}
\label{fig: motivations}
%\end{figure*}
%
\vspace{25pt}
%
%\begin{figure*}[th!b]
\centering
\includegraphics[width=\linewidth, height=5cm]{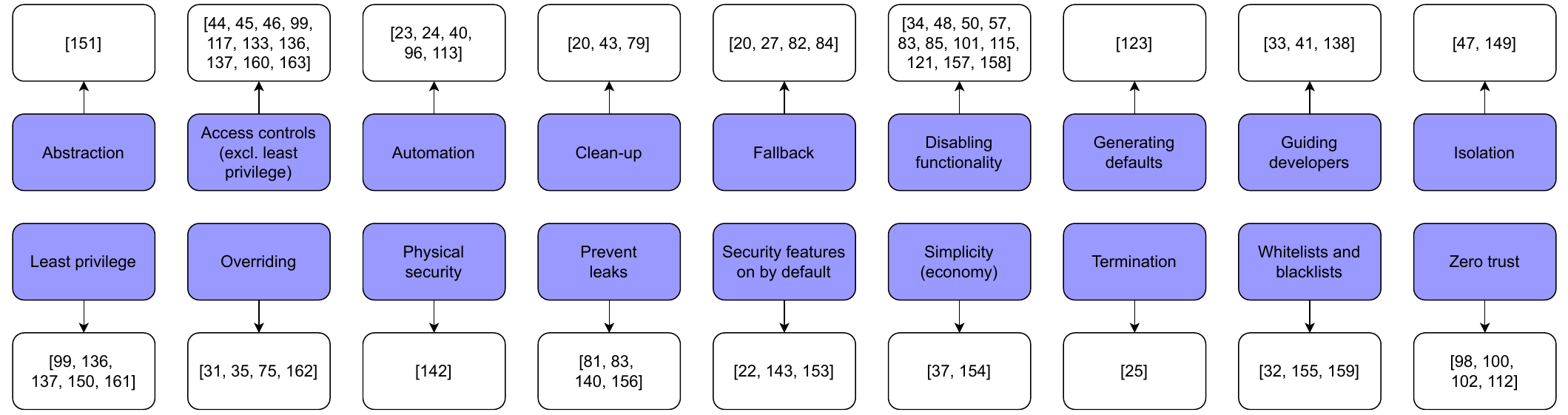}
\caption{Design Principles}
\label{fig: principles}
%\end{figure*}
%
\vspace{25pt}
%
%\begin{figure*}[th!b]
\centering
\includegraphics[width=\linewidth, height=1cm]{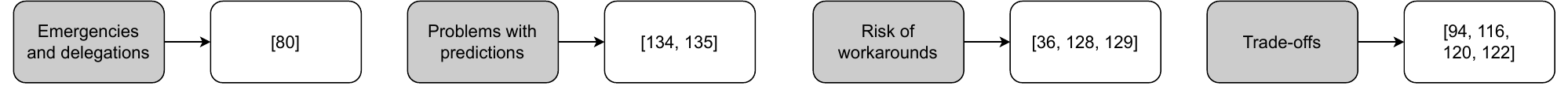}
\caption{Problems}
\label{fig: problems}
\end{figure*}

The security issues in the IoT domain translate also into network protocols. In
particular, numerous publications have analyzed and tried to improve the
so-called MQTT protocol, which does not encrypt traffic by default. In this
regard, the QUIC protocol has been seen as a promising
alternative~\cite{ieee92secure}. However, it should be noted that the issues
with protocols go beyond the IoT and related domains; among other things, the
same-origin policy for the world wide web has been seen to violate the design
paradigm too~\cite{ieee85secure}. In any case, cryptography is also a good
example otherwise; among other things, some publications have analyzed insecure
use of cryptographic libraries; hence, secure defaults should be provided in
these.

Insecure defaults are indirectly reflected also in observations that some
cryptographic libraries are prone to side-channel
attacks~\cite{ieee66secure}. Regarding other common protocols: while the
transport layer protocol (TLS) is the \textit{de facto} one for the today's
world wide web, electronic mail is still unencrypted by default, which has
motivated some to try to improve the situation~\cite{ieee71secure}. Another
example would be nowadays popular end-to-end encryption; people seem to lack
knowledge about it and do not trust it~\cite{ieee84secure}. As always with
cryptography, a point has also been raised that encryption of IoT and other
devices may cause obstacles for forensics and law enforcement
investigations~\cite{ieee60secure}. A further point is that the antonyms have
not motivated only research on IoT, protocols, and cryptography. A good example
would be the operating system domain; therein, some virtualization solutions
have been seen to unnecessarily expose too many system
calls~\cite{ieee87secure}. Another example would be possibly insecure push
notifications used in smartphones by default~\cite{ieee62secure}. All these
examples can be seen to belong to the domain of insecure and unsafe defaults,
the antonyms of the design paradigm.

As can be also seen from Fig.~\ref{fig: motivations}, there have been also other
technology-oriented motivations. Problems with existing access control
implementations are among these. Among other things, some mandatory access
control (MAC) implementations have been seen to violate the paradigm because
programs without a security policy run unconfined by
default~\cite{acm20safe}. Misuse of access authorizations has also been seen as
a more general problem in otherwise secure systems due to imprudent choice of
default privileges~\cite{ieee75secure}. At the same time, some publications have
designed and argued for a middle-ground between a totally open and a strictly
closed system~\cite{ieee79secure}. This point serves to underline that there are
often trade-offs involved, as will be also elaborated later on in a different context.

Three publications used system monitoring, dynamic behavior of networks, and a
risk of a single point of failure as motivations. Remotely conducted monitoring
was seen to provide means to detect misconfigured security
settings~\cite{ieee15secure}, traffic locality and dynamics exhibited were
argued to contradict often coarse-grained and static default network
configurations~\cite{ieee14safe}, and a dynamic conflict arbiter was designed to
prevent single points of failure~\cite{ieee21safe}. It can be noted that in the
non-sampled literature single points of failure are related to the classical
security design principle of defense in depth~\cite{vanOorschot21}. There was
also an odd category labeled as opt-in. For instance, an argument was raised
against the C language in that its use ``rely too much on disciplined opt-in;
there is still no safe default'' \cite[p.~243]{acm15safe}. The category is
closely related to the non-technical motivations labeled as human factors
in~Fig.~\ref{fig: motivations}.

On the human side of things, one of the primary problems---if not the primary
problem---is seen to originate from human behavior and incentives of humans,
whether end-users or developers. The incentives are often perceived as
non-rational in the literature; people allegedly tend to stick with unsafe or
insecure default settings due to various cognitive biases and other
psychological reasons.

Among other things, they may lack knowledge, they may defer changes to settings
due to cognitive biases, including a bias that a default setting conveys
information about what is reasonable, or they may lack technical skills to make
changes~\cite{acm18safe}. There is also a so-called status quo bias; people
allegedly fear that changing security configurations break existing
functionality~\cite{acm18safe, ieee7safe}. Though, the incentives or biases may
also be rational in a sense; people allegedly stick to default security features
in operating systems because they are provided free of
charge~\cite{ieee44secure}. Whatever the actual reasons might be,
some~\cite{acm11secure} or most~\cite{ieee67secure} people never modify default
configurations. Passwords are a good example. For instance, one study found that
``an outstanding 83.2\% of the students have not changed their passwords''
\cite[p.~1431]{ieee4secure}. However, similar claims do not affect only
end-users. Also software developers tend to use default security features
offered by a given provider~\cite{ieee12secure}. Furthermore, there is a
phishing-related angle to this reasoning; even users who follow good security
practices may opt for insecure solutions when presented a choice during their
work flows~\cite{ieee88secure}. There are a couple of problems in these claims
and the associated reasoning about human factors related to default~settings.

The first problem is that the literature sampled contains also contrary
claims. For instance, one study found that default configurations for local area
wireless networks are frequently altered, hinting about a proactive stance to
security \cite{ieee22safe}. The second reason follows: with a couple of surveys
and experiments notwithstanding, much of the reasoning has been speculative. In
other words, most of the claims raised in the literature about human behavior
are not backed by robust and systematized empirical~evidence. Despite these
problems, the human factors have expectedly motivated also usability and user
interface research---after all, also Saltzer and Schroeder perceived ease of use
as essential, connecting it to the psychological acceptability of security
mechanisms (see Table~\ref{tab: saltzer}). For instance, secure default settings
were seen in one paper as being particularly important for the
elderly~\cite{ieee2secure}. However, the literature sampled contains also
critical perspectives; among other things, limited customization was seen as a
problem~\cite{ieee31secure}. Indeed, in case the defaults are insecure, a lack
of customization options may even imply that it is impossibly to remedy the
insecurity by default.

\subsection{Design Principles}

The literature has presented numerous security design principles, some of which
are relatively straightforward security design principles and some others more
general ideas about good security practices to adopt. All of these are on the
technical side; none of the papers sampled provided feasible and
  plausible ideas on how the problems with human factors could be resolved,
provided that generic points raised about improved education and security
awareness are not counted. On these notes, the design principles are
summarized in Fig.~\ref{fig: principles}.

A few papers have operated with high-level design principles. These include the
Saltzer's and Schroeder's \textit{economy} principle. For instance, ``the recent
emphasis on simplifying APIs (and choosing secure defaults) has provided
improvement; we endorse continuing in this direction''
\cite[p.~168]{ieee63secure}. The economy principle has also been correlated with
the psychological acceptability principle in the literature; ``designs should be
as simple as possible and provide users with a secure default
path-of-least-resistance to complete authentication''
\cite[p.~203]{acm10secure}. The notion about path-of-least-resistance is also
known as a user-buy-in security design principle in the non-sampled
literature~\cite{vanOorschot21}. In addition to the economy principle, a~general
\textit{abstraction} design principle and a principle of \textit{isolation} can
be seen to belong to the category of high-level design principles. In terms of
the former, abstracting security features has been argued to reduce a likelihood
that developers will misuse the features, possibly introducing vulnerabilities
along the way~\cite{acm17safe}. Then, isolation has been used to restrict
computer communication between entities~\cite{ieee13secure}, and more generally,
``an isolation-by-default policy results in a significant increase in user
security'' \cite[p.~1342]{ieee36secure}. It should be remarked that isolation as
a default is again nothing new as such; in the non-sampled security literature
it is essentially also behind the compartmentalization design
principle~\cite{vanOorschot21}. Regardless of a terminology, in the literature
reviewed isolation has gained a newfound interest due to the emergence
of virtualization, cloud computing, and associated technologies and services.

As already noted, quite a few publications have operated with \textit{access
  controls} and the \textit{least privilege} principle considered already by
Saltzer and Schroeder. The works in this domain often use apt slogans to
motivate their designs. ``Deny by default'' \cite[p.~509]{acm3safe} is a good
example. Although the basic idea remains the same, the slogans vary from a
context to another. For instance, publications dealing with input and output
operations have justified their designs by security policy notions such as a
``fail-safe default of no write access'' \cite[p.~119]{acm19safe} or a default
``to automatically deny the read-write'' \cite[p.~260]{ieee11safe}. These access
control slogans and security policy dictates align with a principle of
\textit{blacklists and whitelists}. Before continuing, it should be noted that
the former violates the Saltzer's and Schroeder's original fail-safe default
principle, whereas the latter conforms with it. Among other things, whitelists
have been used to restrict parameters supplied to cryptographic
libraries~\cite{acm8safe}, whereas ``billions of devices benefit from the
blacklist enabled by default'' \cite[p.~4345]{ieee3safe} in web browsers. This
quotation demonstrates that it is not always possible to fulfill the ideal of
denying by default; there are billions (or more) of legitimate websites, and
thus no one can curate a whitelist in practice.

A principle of \textit{disabling functionality} is prominent in the
literature. The explanation is partially related to the overall insecurity in
the IoT domain. A few illuminating quotations help to again elaborate this
principle. ``Making communication default-off has tangible security benefits''
\cite[p.~117]{ieee9safe}. This ``off-by-default approach''
\cite[p.~197]{ieee4safe} is commonly endorsed in the computer network research
domain; all communication is denied by default~\cite[p.~321]{ieee14secure} and
computer networks have been designed so that they ``only admit the expected
traffic, by default treating the rest as \textit{unwanted} traffic''
\cite[p.~1]{ieee57secure}. In fact, recent ``proposals for capabilities-based
networks have provided some ideas on the fundamental shift in the design
philosophy of networks by moving from the Internets `on-by-default' principle to
an `off-by-default' assumption'' \cite[p.~1931]{ieee16secure}. However, the
principle generalizes to other domains as well. For instance, web security has
been argued to improve from ``disabling the behavior by default in web
browsers'' \cite[p.~673]{ieee15safe}, including with respect to browser
plugins~\cite{ieee1secure}. Furthermore, the off-by-default dictum has been
adopted as a general argument about delivering software products; they ``should
really be distributed with access disabled until administrators explicitly
customize the access control policy and mechanisms for their organization''
\cite[p.~10]{acm2secure}. Finally, the principle also aligns in the literature
with the economy principle; a design ``aims for true minimality in the sense
that nothing should be included by default that the service does not explicitly
need'' \cite[p.~251]{ieee53secure}. It is not difficult to come up with other
slogans aligning with the principle---it could be described also with phrasing
such as ``disable by default, then enable'', ``close by default, then open'',
and so forth and so on.

The off-by-default dictum should not be generalized to everything; the
literature also emphasizes a principle of turning \textit{security features on
  by default}. In addition to technical designs and their implementations,
including those mandating an execution of specific
instructions~\cite{ieee17safe} and those related to enabling sanitization
routines in web applications~\cite{acm8secure}, the principle aligns with the
human factor side. The incentives discussed earlier are visible in observations
that study participants ``were happy to use two-factor authentication as a
secure default that is set up at registration, while they would not go through
the steps to set it up afterwards'' \cite[p.~12]{acm9secure}. Thus, also this
principle aligns with the psychological acceptability principle; people should
be encouraged to adopt good security practices, whether via education or user
interface designs.

There is also a principle related to \textit{overriding} functionality enabled
by default. For instance, a design hooked calls to existing APIs in order to
enforce validation of host names and certificates~\cite{acm13safe}. Another good
example involves computer networking; a study designed a system via which ``the
default credential is \textit{exchanged} for a credential that is suitable for
the service in question'' \cite[p.~150]{ieee86secure}. Though, an argument can
also be raised that overriding things should not be necessary in the ideal world
because it indicates that something is non-optimal in a design that is being
overridden.

To some extent, the already noted \textit{clean-up} design principle aligns with
a principle to \textit{prevent leaks} of different kinds. Such information or
data leaks vary from a context to another. A good example would be enforcing
``best practices when enabling logs, ensuring the server administrator really
intended to expose this information'' \cite[p.~4]{acm7secure}. Another example
would be plugging leaks in honeypots in order to improve deception and prevent
detection by adversaries~\cite{ieee26secure}. In the general, non-sampled
literature the detection techniques are known as anti-honeypot and
anti-introspection methods~\cite{Uitto17}. As could be expected, the leak
prevention principle has also been adopted to protect the privacy of users by
default~\cite{ieee13safe, ieee50secure}. Thus, leaks are a good example on how
security design principles sometimes extend to other domains as well.

Then, there are design principles related to \textit{automation} and
\textit{generating defaults}. Regarding the former, examples include automation
of network security configurations~\cite{ieee33secure}, providing personalized
security settings~\cite{ieee31secure}, automation of virtual machine
configurations~\cite{ieee48secure}, and automatic provision of certificates to
end-users~\cite{ieee91secure}. Regarding the default generating principle, a
paper presented a technique to ensure a secure default option by generating
strong passwords automatically for users~\cite{acm6secure}. While a full
automation of security configurations may be challenging and thus also risky,
the latter idea would seem sensible for also fixing some of the basic issues
with IoT devices; user names and strong passwords could be generated at a
factory and perhaps printed to stickers placed at the bottoms of all IoT devices
manufactured and shipped.

There is a further design principle related to explicitly \textit{guiding
  developers} about security. For instance, a paper designed a fail-safe default
that forces developers to check and validate their
configurations~\cite{acm9safe}. Another paper dealing with databases designed a
solution ``forcing security modellers to think about any cycles in a schema, and
add explicit constraints to weaken the schema only where they believe this is
safe'' \cite[p.~113]{acm21safe}. This principle aligns with the earlier points
about secure defaults for cryptographic libraries, further strengthening the
defaults with explicit constraints.

Furthermore, there are design principles about robust \textit{fallback}
and \textit{termination}. Some of these principles are on the safety side and
the domain of cyber-physical systems. For instance, a custom solution for
unmanned aerial vehicles (UAVs) was compared against a ``default mitigation
strategy'' involving opening a ``parachute when a failure is
detected''~\cite{ieee12safe}. Another paper operating in the UAV domain
discussed a fail-safe default of either landing or returning to
home~\cite{ieee18safe}. That said, sometimes there is no other option but to
gracefully terminate an execution; ``stopping locomotion when an error is
detected is a safe default action'' \cite[p.~150]{acm10safe} in some robotics
applications. Closer to the domain of software and systems, the fallback design
principle often involves reverting to a last known good or valid state in case
of a malfunction~\cite{ieee1safe}. Some of the fallback designs also correlate
with the other principles outlined. Among these is isolation; ``monitor the
system and rapidly adapt when conditions change, falling back to strict
isolation as the safe default'' \cite[p.~351]{acm11safe}. Analogously, a
``system responded by disabling all network ports and defaulted to the
safe-operation mode established previously'' \cite[p.~1]{ieee6safe}. While these
and other fallback and termination designs were not considered by Saltzer and
Schroeder, they too can be seen to belong to the principle of fail-safe
defaults.

Finally, in addition to an outlying paper implicitly dealing with a principle of
\textit{physical security}, three papers built upon the emerging \textit{zero
  trust} principle. Its motivation can be located also from Fig.~\ref{fig:
  motivations}, which includes a paper discussing the many problems that emerged
from an ill-conveyed historical principle of trusting by default. In general,
the zero trust principle of never trusting and always verifying is often taken
to imply rigorous monitoring and fine-grained access controls because an
underlying assumption is that a part of a larger computing infrastructure has
already been compromised. To this end, all ``requests are not trusted to access
the system networks by default unless passing the anomaly detection''
\cite[p.~4020]{ieee38secure}. Analogously, a paper's design stated that ``no
user or system can be trusted by default'' \cite[p.~79]{ieee61secure}. In fact,
by ``default, our allocation policy considers all legitimate users as
attackers'' \cite[p.~440]{ieee47secure}. These and other related
characterizations of the zero trust principle further strengthen the classical
principles of fail-safe defaults and complete mediation initiated by Saltzer and
Schroeder.

\subsection{Problems}

The literature reviewed has also discussed some notable problems, obstacles, and
limitations in adopting the design paradigm. Four such problems were identified
during the literature review (see Fig.~\ref{fig: problems}). Thus, to begin
with, first, a paper operating in the healthcare domain noted that sometimes
access controls must be circumvented in order to save lives of patients, further
emphasizing that fully locked systems may prevent necessary delegations to some
trusted parties in emergency situations~\cite{ieee10secure}. In the non-sampled
literature this problem is sometimes known as breaking the
glass~\cite{Brucker09}. The second important problem is related to predictions
about the future.

In particular, it was argued that design ``principles are based on tacit
assumptions that were true in the past but possibly false now''
\cite[p.~92]{ieee2safe}. Analogously, ``today's secure default becomes
tomorrow's vulnerability'' and ``we question whether it is possible to design,
before use, a secure default that can anticipate every possible use''
\cite[p.~54]{acm4secure}. These arguments signify the earlier point about a need
of empirical research. Further research is particularly needed to better
understand whether and how the design paradigm adapts when systems, software,
and networks evolve through time. In addition, the quotations underline a need
to study how threat modeling could be improved, such that anticipation could be
better---even in case perfect anticipation is reasonably taken as an impossible
task to fulfill in practice.

The third area of problems was seen to originate from a risk of workarounds
developed by software developers in particular. For instance, a paper operating
in the domain of programming languages noted that ``programmers will often
choose a `safe' default data type, often double floating-point precision,
whether or not it is appropriate'' \cite[p.~8:2]{acm16safe}. The problems were
seen to also relate to the economy design principle. Among other things, the
efforts to simplify often ``remove choices available to developers, which leads
to additional mistakes when programmers develop workarounds''
\cite[p.~182]{acm5secure}. Regarding the mistakes involved, which can be severe,
a paper found that developers introduced vulnerabilities by disabling security
measures imposed and even broke cryptography~\cite{ieee5secure}. Against such
results, it might be contemplated whether the guiding developers design
principle should involve softer motivating forms, such as those known with a
term nudging.

Fourth, the literature also discussed different trade-offs behind the design
paradigm. As could be expected, particularly a trade-off between security,
including particularly in terms of encryption, and performance was raised in the
literature~\cite{ieee28secure, ieee64secure}. The trade-off context could be
extended to cover also the human factors; the relation between usability and
security has been debated ever since the work of Saltzer, Schroeder, and other
early titans. Having said that, the trade-offs as well as the workaround problem
could be argued to involve implementations and not the design paradigm in
itself. If the paradigm would be properly implemented to begin with, it could be
argued the problems would not be present.

\section{Discussion}\label{sec: discussion}

In what follows, a summary of the review is first presented after which a couple
of limitations are noted. The final subsection points out four areas for
further~research.

\subsection{Conclusion}

This systematic mapping study and a scoping review addressed the security design
paradigm of secure and safe defaults. In total, $n = 148$ mostly high-quality
peer reviewed papers were reviewed. The primary conclusions can be summarized
with six concise points.

First, it is evident that the domains in which the paradigm has been discussed
and applied have considerably expanded over the years. The Saltzer's and
Schroeder's original context of access controls and operating systems is still
present, but the paradigm has been applied, discussed, and extended in numerous
other domains as well. Regarding the original context, virtualization, cloud
computing, and related technologies and infrastructures have brought a
revitalized interest in the paradigm and the associated security design
principles, including the isolation design principle discussed in a couple of
papers.

Second, the computer networking domain, including IoT devices, has discussed the
paradigm particularly actively. The reason is simple: IoT devices are widely
seen to violate the paradigm. Whether it is default or otherwise insecure
passwords or a lack of update mechanisms, the IoT domain has been seen to
operate with the paradigm's antonyms, insecure and unsafe defaults. The point is
important because the majority of papers reviewed have motivated themselves
through a negation of the paradigm. The computer networking and IoT domains are
further important to emphasize because it seems a new ``off by default'' design
principle seems to be \text{emerging---or} has already emerged---therein. It is
also related to a principle about disabling unneeded or unnecessary
functionality. It remains to be seen whether these principles help at remedying
some of the notable issues plaguing IoT~devices.

Third, regarding the contextual domains, it is important to emphasize that the
paradigm is not only about security but also about safety. Cyber-physical
systems are a good example in this regard. The questions about how to fail
safely and what are safe defaults are quite different in this contextual domain
compared to the other, more traditional computing domains.

Fourth, the motivations behind the papers reviewed can be reasonably well
categorized into technical problems and those dealing with human behavior. This
categorization aligns with existing argumentative reviews~\cite{Ryan23} with an
exception that organizational aspects are entirely missing from the literature
sample gathered and reviewed. Thus, without further reviews, nothing can be said
about safe and secure defaults in terms organizational security, among other
things. However, with respect to the noted argumentative review, it is important
to emphasize that the literature has considered human behavior both in terms of
end-users and developers or professionals. Having said that, there are also some
notable problems in this regard, as soon discussed. It is also worth pointing
out that the absence of organizational security might well be due to the
sampling from the ACM's and IEEE's electronic libraries alone. In other words,
particularly social sciences are \text{missing---as} again soon pointed out,
they too have interesting things to say about secure and safe~defaults.

Fifth, the design paradigm has been discussed and developed with many security
design principles, some of which can be traced to the work of Saltzer and
Schroeder, but some of which are relatively new. Regarding the newer principles,
it is worth noting that the ``off by default'' design principle is matched by an
``on by default'' principle in terms of some or all security features
available. In addition, principles related to security configuration automation,
clean-up routines, overriding existing functionality, and leak prevention can be
mentioned as new design principles. Many papers have also designed different
fallback solutions. The emerging zero trust design principle is also visible in
the literature sample reviewed.

Sixth and last, the literature reviewed has also discussed different
problems developers and others have faced when trying to apply the
design paradigm. These problems include traditional concerns, such as a
relation between security and performance, but also interesting points have been
raised about human behavior. Among other things, some have argued that providing
secure defaults increase a likelihood that developers will adopt potentially
dangerous workarounds. Another important point raised in the literature is about
predicting the future. In particular, an argument has been raised in the
literature that today's defaults may be tomorrow's vulnerabilities. The point is
related to the ``unhelpful assumptions'' discussed
recently~\cite{Ryan23}. However, a critical argument can be raised that
argumentation with such assumptions may be prone to the classical false balance
bias---it seems counterproductive to wholesale abandon the paradigm just because
there may be some unintended consequences.

\subsection{Limitations}

A couple of notable limitations should be acknowledged. The first limitation is
that the review was conducted by a single author. This limitation affects both
the literature sampling protocol and the qualitative thematic analysis discussed
in Section~\ref{sec: methodology}. Therefore, it is also impossible to provide
inter-rater reliability measures commonly used for systematic literature reviews
and mapping studies~\cite{Zolduoarrati23}. However, it remains unclear and
debatable how the limitation should be addressed in practice. The reason for
this claim originates from the second limitation; it also remains unclear how
the $n = 148$ papers reviewed generalize toward an unknown theoretical
population of all papers discussing the paradigm. The two limitations are linked
together due to the feasibility constraint noted earlier. Covering even more
papers would quickly face the constraint, which would be also encountered by
involving multiple authors assessing the papers and the thematic
constructs. This point is not unique to the review at hand.

As has recently been argued, systematic literature reviews and systematic
mapping studies have become increasingly bloated and increasingly infeasible to
conduct~\cite{dosSantos25}. To some extent, the reason is simple: the volume of
academic literature grows year by year, making the feasibility problem more
severe step by step. To this end, alternative reviewing techniques, such as
scoping reviews, have been proposed. There is also a discussion about using
artificial intelligence tools for literature reviews~\cite{Kraus24,
  Schmitt25}. The review presented supports this line of reasoning---feasibility
is a constraint also in the cyber security domain, and further discussion is
needed on whether and how it could be addressed in the future.

\subsection{Further Research}

What about knowledge gaps and further research? Four points can be raised in
this regard. The first point is that empirical research has mostly been
absent. Although empirical research may be difficult to initiate in the domains
of systems and network engineering, particularly the assumptions about human
behavior would need rigorous empirical assessments. As it stands, many of the
assumptions and theorizations presented in the literature more or less fall into
a category of folklore---a criticism that is hardly unique to the literature
reviewed~\cite{Fernandez19}. This folklore category is related to the unhelpful assumptions noted.

The second point is that the security engineering domain would benefit from a
systematic catalog of security design principles. Such a catalog might be
designed by following the existing summaries on the bodies of existing knowledge
in some computing disciplines~\cite{IEEE04}. It would be also useful for both
researchers and practitioners to continue with the work initiated for
translating security design principles into more concrete security design
patterns~\cite{Schumacher03, vandenBerghe17}. These points are
important because the review also indicated that many security design principles
are discussed with different, often overlapping concepts. The Saltzer's and
Schroeder's principle of psychological acceptability is a good example; in more
recent usability and user interface engineering research it aligns with the
principle of least surprise~\cite{Horcher19}. As was discussed, the same point
applies to other security design principles. For instance, the high-level
software design principle of ``building security in''~\cite{McGraw04} did not
appear in the literature reviewed, although it too can be seen to be a part of
the paradigm.

The third point is that a full coverage of insecure default configurations has
not been covered in the context of the design paradigm. Although the IoT domain
has justifiably raised alarm bells in this regard, and there are works
discussing configuration vulnerabilities already in the early 2010s---if not
earlier, some notable recent domains are absent in the literature sampled. A
good example would be vulnerabilities involving cloud computing and related
configurations~\cite{Continella18, Rahman23}. This example serves to underline
that configuration vulnerabilities may affect practically all computing
domains. Secure defaults serve as a starting point for mitigating such
vulnerabilities.

The fourth and last point is that research on regulations is lacking behind. In
this regard, a paper reviewed raised an argument that ``policymakers should
ensure the default setting be set to enable security''
\cite[p.~270]{acm18safe}. Although regulations constitute a large and complex
research domain of their own, the Cyber Resilience Act (CRA) recently agreed
upon in the European Union (EU) is a good example with respect to the design
paradigm.\footnote{~Regulation (EU) 2024/2847.} It imposes many new security
requirements for most information technology products, whether software or
hardware based. Among other things, obligations are placed upon both commercial
vendors and open source software projects in terms of vulnerability disclosure
and supply-chain security management~\cite{Ruohonen24JSS,
  Ruohonen24IFIPSEC}. For the present purposes, it is important to emphasize
that the so-called essential cyber security requirements upon which compliant
products can be placed in the future to the EU's internal market include a
clause that the products should be distributed ``with a secure by default
configuration''.\footnote{~Paragraph 2(b) in Annex I.} This point serves to
emphasize that the design paradigm is well-recognized also on the side of
policy-makers and regulators, either explicitly or implicitly. The point can be
also connected to the earlier remark that social science research is missing
from the review. Without trying to delve deeper into the probably large amount
of research on these sciences about default settings, an example can be
mentioned about a relation between the currently prevailing opt-out defaults and
privacy, misinformation, manipulation, and related online
phenomena~\cite{Grill22, Ruohonen24FM}. Such a relation between default settings
and online phenomena serves well to end the mapping and scoping review with a
remark that the design paradigm extends well-beyond technical work related to
security and safety by default.

\balance
\bibliographystyle{ieeetr}
%\bibliography{secdefaults}

\end{document}